\documentclass[11pt]{article}
\usepackage[utf8]{inputenc}
\usepackage{graphicx}
\usepackage[frozencache,cachedir=.]{minted}
\usepackage[linesnumbered,ruled,noend,vlined]{algorithm2e}
\usepackage{hyperref}    
\usepackage{enumerate}
\usepackage{booktabs}

\usepackage{boxedminipage}
\newenvironment{result}
{\smallskip
\noindent
\let\emph=\textbf
\begin{boxedminipage}{\columnwidth}\em}
{\end{boxedminipage}
}

\usepackage{geometry}
\title{Exploring API Behaviours Through Generated Examples}

\author{Stefan Karlsson\textsuperscript{1,2}, John Hughes\textsuperscript{3,4} Robbert Jongeling\textsuperscript{2}, Adnan \v{C}au\v{s}evi\'{c}\textsuperscript{1}, Daniel Sundmark\textsuperscript{2}\\ \textsuperscript{1}ABB AB, Västerås, Sweden. \\ adnan.causevic@se.abb.com  \\ \textsuperscript{2} Mälardalen University, Västerås, Sweden. \\ \{stefan.l.karlsson, daniel.sundmark, robbert.jongeling\}@mdu.se\\ \textsuperscript{3} Chalmers University of Technology, Gothenburg, Sweden. \\ rjmh@chalmers.se \\ \textsuperscript{4} Quviq AB, Gothenburg, Sweden. }
\date{}

\begin{document}

\maketitle

\begin{abstract}
Understanding the behaviour of a system's API can be hard. Giving users access to \textit{relevant} examples of how an API behaves has been shown to make this easier for them. In addition, such examples can be used to verify expected behaviour or identify unwanted behaviours.

Methods for automatically generating examples have existed for a long time. However, state-of-the-art methods rely on either white-box information, such as source code, or on formal specifications of the system behaviour. But what if you do not have access to either? e.g., when interacting with a third-party API. 

In this paper, we present an approach to automatically generate relevant examples of behaviours of an API, without requiring either source code or a formal specification of behaviour. 

Evaluation on an industry-grade REST API shows that our method can produce small and relevant examples that can help engineers to understand the system under exploration.
\end{abstract}

{\bf Keywords}: Property-based testing, Examples, Automated testing, API testing, REST

\begin{figure*}
    \centering
    \includegraphics[width=\textwidth]{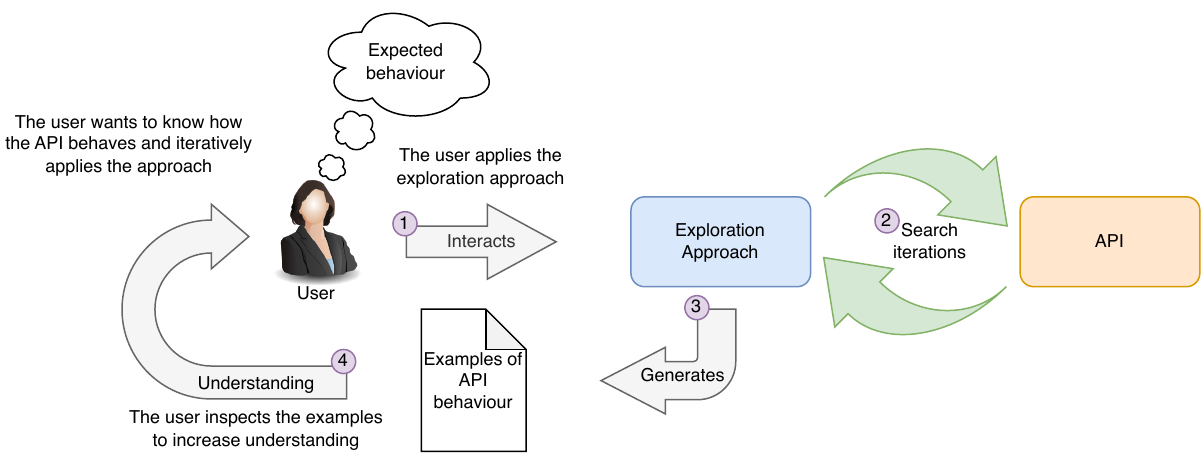}
    \caption{The proposed approach can be used in an interactive exploration process, where a user---based on the output---refines the understanding of an API}
    \label{fig:interactive-process}
\end{figure*}

\section{Introduction}\label{sec:introduction}

Understanding and verifying the behaviour of an API can be a difficult task. One way of alleviating some of this burden is through access to examples of the API's behaviour~\cite{Robillard-What-Makes-APIs-Hard-to-Learn?-Answers-from-Developers-2009, Robillard-A-field-study-of-API-learning-obstacles-2011, McLellan-Building-more-usable-APIs-1998, Nykaza-What-Programmers-Really-Want:-Results-of-a-Needs-Assessment-for-SDK-Documentation-2002, Shull-Investigating-reading-techniques-for-object-oriented-framework-learning-2000, Novick-What-Users-Say-They-Want-in-Documentation-2006}. Generated examples can significantly aid users in understanding the behaviour of an API~\cite{Gerdes-Understanding-Formal-Specifications-through-Good-Examples-2018} and many approaches have been proposed to automatically produce them~\cite{Buse-Synthesizing-API-usage-examples-2012, Barnaby-Exempla-Gratis-(E.G.)-Code-Examples-for-Free-2020, Gu-CodeKernal-2019, Kim-Adding-Examples-into-Java-Documents-2009, Mar-Recommending-Proper-API-Code-Examples-for-Documentation-Purpose-2011, Montandon-Documenting-APIs-with-examples-Lessons-learned-with-the-APIMiner-platform-2013, Holmes-Approximate-Structural-Context-Matching:-An-Approach-to-Recommend-Relevant-Examples-2006, Moreno-How-Can-I-Use-This-Method-2015}. A common theme for these approaches is that they require access to white-box information such as existing usage examples from source code. In contrast, Gerdes et al. proposed a black-box approach~\cite{Gerdes-Understanding-Formal-Specifications-through-Good-Examples-2018}. However, it required not only the implementation itself, but also a formal specification of the code, from which tests could be generated. The formal specification played a key role in selecting examples of ``interesting'' behaviour.

What if we do not have a specification? Can we still generate tests to {\em explore the behaviour of the system}, and among them select interesting examples from which the user can gain a new understanding of the system's behaviour? Can we automate the behaviour of ``playing with the system'' to understand parts of what it does? Gaining such understanding of a system is vital for: end-users in using an API to successfully interact with the system, testers verifying the actual behaviour of a system, and developers making changes to the system, wanting to ensure the correct behaviour has been altered.

In our work, we consider this situation and generate examples of the system's behaviour without requiring the source code or a formal specification. As illustrated in Figure~\ref{fig:interactive-process}, an example generating approach, as we propose, fits in an interactive workflow where understanding of an API is refined over time.

To be useful, we must consider the following three requirements for generating examples; (i) generated examples should illustrate interesting and different behaviours, i.e., how different API operations interact~\cite{Robillard-A-field-study-of-API-learning-obstacles-2011}. Moreover, to avoid overwhelming the user, (ii) the number of examples should be small and (iii) each example should be minimal~\cite{Robillard-A-field-study-of-API-learning-obstacles-2011}. Indeed, a generated example consisting of a random sequence of API operations would not be helpful in understanding a system's behaviour. Therefore, we focus on generating \emph{relevant} examples.

Potential uses of generated examples include helping users and developers understand the behaviour of an API. Users could be, e.g., end-users who want to understand its functionality better, or developers who make changes to the codebase that the API exposes. For the latter type of user, generated examples of the actual behaviour could then show both expected pre-existing behaviours and unexpected new behaviours, whereas an existing test suite would be limited to finding potential regressions in existing behaviours, but would not uncover new, potentially unwanted, behaviours.

Another potential use case is in systems that use a microservice architecture \cite{Fowler-Microservices}, where the system is composed of many services (sometimes hundreds). For a developer implementing an API for such services, generated examples can serve as a basis for usage documentation, speeding up the documentation process. In addition, examples of how operations interact could yield documentation in a tutorial fashion, rather than a flat list of API operations. In this way, the generated examples would complement a typical API reference documentation. Example interactions would alleviate the need for the user to have a mental model of the state of the system and how the operations interact when reading the API documentation.

\begin{figure*}
    \centering
    \includegraphics[width=\textwidth]{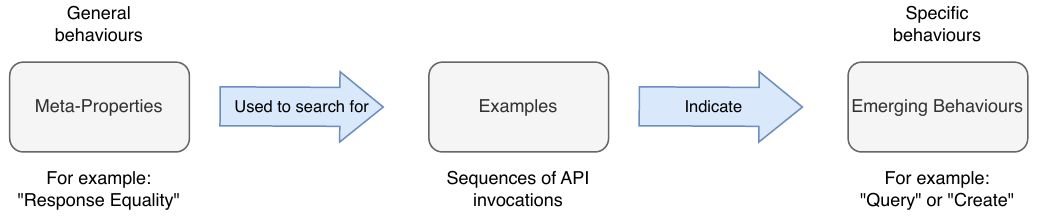}
    \caption{Meta-Properties define general behaviours. Examples are produced from the meta-properties. A set of examples, or the lack of, indicate specific emergent behaviours exposed by an API.}
    \label{fig:concepts}
\end{figure*}

In this paper, we present an approach for generating relevant examples of API behaviours. Our approach generates examples as sequences of operations invoking the API. The examples are generated by automatically searching for predefined commonly occurring categories of general behaviour that we expect many systems to display instances of. We call these general behaviours ``meta-properties''. A set of generated examples of general behaviours can indicate specific behaviours of an API. These indicated specific behaviours emerge from the generated examples. An example of such a specific behaviour is that of a query-operation---an operation that has the general behaviour of not changing the state of the system. Figure~\ref{fig:concepts} shows how these concepts fit together in our proposed approach; meta-properties are used to search for examples of general API behaviours which then indicate specific API behaviours.

Specifically, we make the following contributions.
\begin{itemize}
    \item An approach to automatically generate {\em relevant} examples of actual system-under-test behaviours, as exposed by an API, without the need for a formal specification.
    \item A set of general abstract meta-properties used to categorise behaviours.
\end{itemize}

We evaluate the usefulness of the meta-properties to generate relevant examples on an industry-grade REST API.

The remainder of this paper is structured as follows. Section~\ref{section:relevant-examples} introduces the main concepts of this paper, and defines \emph{relevance}. Section~\ref{section:key-ideas} presents the required key ideas our approach is built upon. In Section~\ref{section:exploring-behaviours}, we describe our approach in detail, which is followed by the definition of a running example in Section~\ref{sec:running-example} and applying our approach to it in Section~\ref{section:an-exploration-example}. Section~\ref{section:evaluation-symbolic-refs} compares the performance of different strategies of parameter selection, which is important to efficiently generate examples.
We find random references to be the best performing means of parameter selection and use that strategy in our evaluation of the approach in Section~\ref{section:evaluation}.
Finally, we present the related work in Section~\ref{sec:related-work} and conclude the paper in Section~\ref{sec:conclusions}.

\section{What is a Relevant General Example?}\label{section:relevant-examples}

We have emphasised that the generated examples should be \emph{relevant}. But what does it mean for an example to be relevant? We agree with previous work that to be ``interesting'' an example should show an interaction between operations, and that all the operations in the example should be essential to its point~\cite{Gerdes-Understanding-Formal-Specifications-through-Good-Examples-2018, Robillard-A-field-study-of-API-learning-obstacles-2011}. 

In this work, we judge an example as relevant if it shows an interaction between the state of the system and a sequence of two or more operations in such a way that a distinct behaviour is demonstrated. For example, a sequence of three operations that query the state of the system, create an entity, and query the state again could demonstrate a state-changing sequence in which the state of the system is enlarged---the system stores a larger number of entities. We exclude sequences of only one operation due to the fact that such a sequence cannot produce an example of interaction between the operation and the state of the system. We would have no general behaviour to relate the operation to, since we do not know how it behaves over multiple invocations or in sequence with other operations. In summary; a relevant example consists of at least two operations and shows a distinct behaviour.

Examples generated from a formal specification have been shown to significantly increase human understanding of the behaviour of an API~\cite{Gerdes-Understanding-Formal-Specifications-through-Good-Examples-2018}. However, this previous work based the example generation on a manually created formal specification of the API, which may not always be available. In this work, instead, we generate examples from \emph{generally} defined API behaviours.
Defining general behaviours, such as we propose, trades losing specificity of the behaviours of a particular API for less required work in specifying the expected behaviours.

Our proposed approach composes these two concepts---relevance and generality---and produces instances of relevant general examples of API behaviours. An example of such a general behaviour is operations that change the state of the system, either by creating, updating or deleting an entity.

\section{Key Ideas}\label{section:key-ideas}

\begin{figure*}
    \centering
    \includegraphics[width=\textwidth]{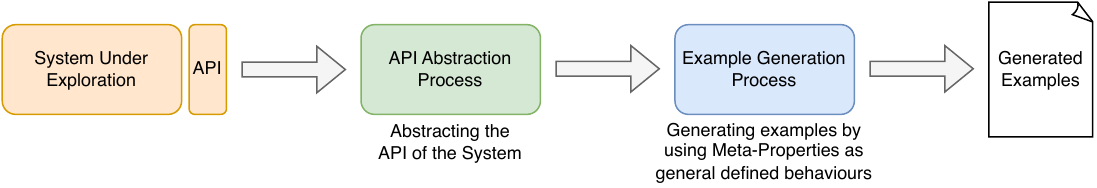}
    \caption{An overview of the key ideas and how they fit in the process of generating examples}
    \label{fig:key-ideas}
\end{figure*}

We propose an approach to automatically generate \textit{relevant} examples of sequences of API calls to help engineers explore the behaviour of their system. We do so without requiring any access to source code or a formal specification of behaviours. Our approach is generic and does not rely on a specific type of system. Consequently, it is not tied to a specific programming language or type of API (such as REST~\cite{fielding-REST-2000} or GraphQL\footnote{\url{https://graphql.org/}}). To enable example generation in the general case, our approach is built on the following key ideas, which are further elaborated in the remainder of this section:
\begin{enumerate}
    \item \textbf{Abstraction of the system API:} To allow interaction with the system without any white-box information and independence from a specific domain, we create an abstract representation of the API of the system.
    \item \textbf{Defining general API behaviours:} In order to generate examples without a formal specification, and to be more generally applicable, we define general API behaviours in what we call \textit{meta-properties}, i.e., properties defining behavioural properties of a system.
\end{enumerate}

 With the use of defined general categories of API behaviours, i.e., \emph{meta-properties}, and an abstract representation of the API, we generate examples that show interactions between the state of the system and a sequence of two or more operations. We do so in such a way that a distinct behaviour is demonstrated, i.e., in a way that the examples are \emph{relevant}. How the key ideas fit into the overall process of generating examples of an API, is shown in Figure~\ref{fig:key-ideas}. These high-level abstract steps are further detailed in Section~\ref{section:exploring-behaviours}, with details included in Figure~\ref{fig:overview}.

\subsection{Abstracting the API of the System}

To enable example generation without requiring a formal specification of behaviours, or other white-box artefacts such as source code, we must abstract the specifics of the system we want to explore. In addition, we only want the user to provide the essential information needed to generate examples.
This essential information must contain (i) what type of API it is (since the way to call the API differs between type of APIs), and (ii) what operations are provided by the API.
The input information is thus limited to what is technically required to call the API.
For example, we do not require any information on how the operations relate, in what order they must be called, or what the dependencies between parameters are. These blanks will be filled by our proposed approach.

By abstracting the specifics of the API, we enable our approach to be used on different APIs of different types. For example, an API might expose the operations of ``query-users'' and ``create-user''. In our approach, these operations are abstracted as ``operation-1'' and ``operation-2''. They do not convey any meaning to the approach --- only their behaviours do. The example generation process uses the abstract representation of the API when searching for examples and is thus agnostic to the specifics of the system providing the API being explored. The approach generates sequences of abstracted operations and assesses if they conform to general behaviours, i.e., no domain information is needed since domain specific behaviours are not assessed. However, as the example generation process uses abstract versions of the actual operations provided by the API---which cannot be used to actually invoke the API---a translation component is required to translate the abstract operation to the actual operation. For example, before actually executing the operation the translation component will then translate the abstract ``operation-1'' back to a concrete API operation, as ``query-users''. This translation can be done once per API type.

\subsection{Meta-properties for defining general API behaviours}
APIs come in many different flavours. To support automatic generation of examples for different APIs, we must base the examples on some general behaviours common to them. We encapsulate these general behaviours in what we call \emph{meta-properties}. A meta-property describes an expected outcome of interacting with a system given a sequence of operations. To adhere to a behaviour defined in a meta-property, constraints can be put on the expected outcome or on the sequence of operations, e.g., the number of operations, or the order of operations given information from a previous example generation result. 
An example of a behaviour that could be captured in a meta-property is: ``a sequence of operations that changes the state of the system''. We assess whether such a meta-property holds or not, based on the observations we make of responses from the API when we execute a generated sequence of operations. In essence, a meta-property defines behaviour in terms of how responses from executing a sequence of generated operations relate.

Meta-properties are central to our approach as they define the behaviours that our exploration searches for. Our approach is extendable in that the process of generating examples is not tied to which meta-properties we use, it is only required that some are defined. Thus, the approach is relevant to many different kinds of APIs, since new meta-properties can be expressed that are relevant to new kinds of APIs. 

Although the exploration process is general, we need to be specific to show the applicability of the proposed approach for some type of API. Therefore, in this paper, we primarily target APIs that manipulate internal entities. Such a behaviour would include  \emph{Creating}, \emph{Querying}, and \emph{Deleting} entities stored in the system. We believe that these basic operations are general and cover many APIs. One type of APIs where this set of operations is commonly found are web-services designed according to a RESTful \cite{fielding-REST-2000} architecture. The services supporting these operations are commonly called CRUD services (Create, Read, Update, Delete). Section \ref{section:exploration-properties} describes the definition of these meta-properties in more detail.

\section{Our Approach for Exploring Behaviours}\label{section:exploring-behaviours}

\begin{figure*}
    \centering
    \includegraphics[width=\textwidth]{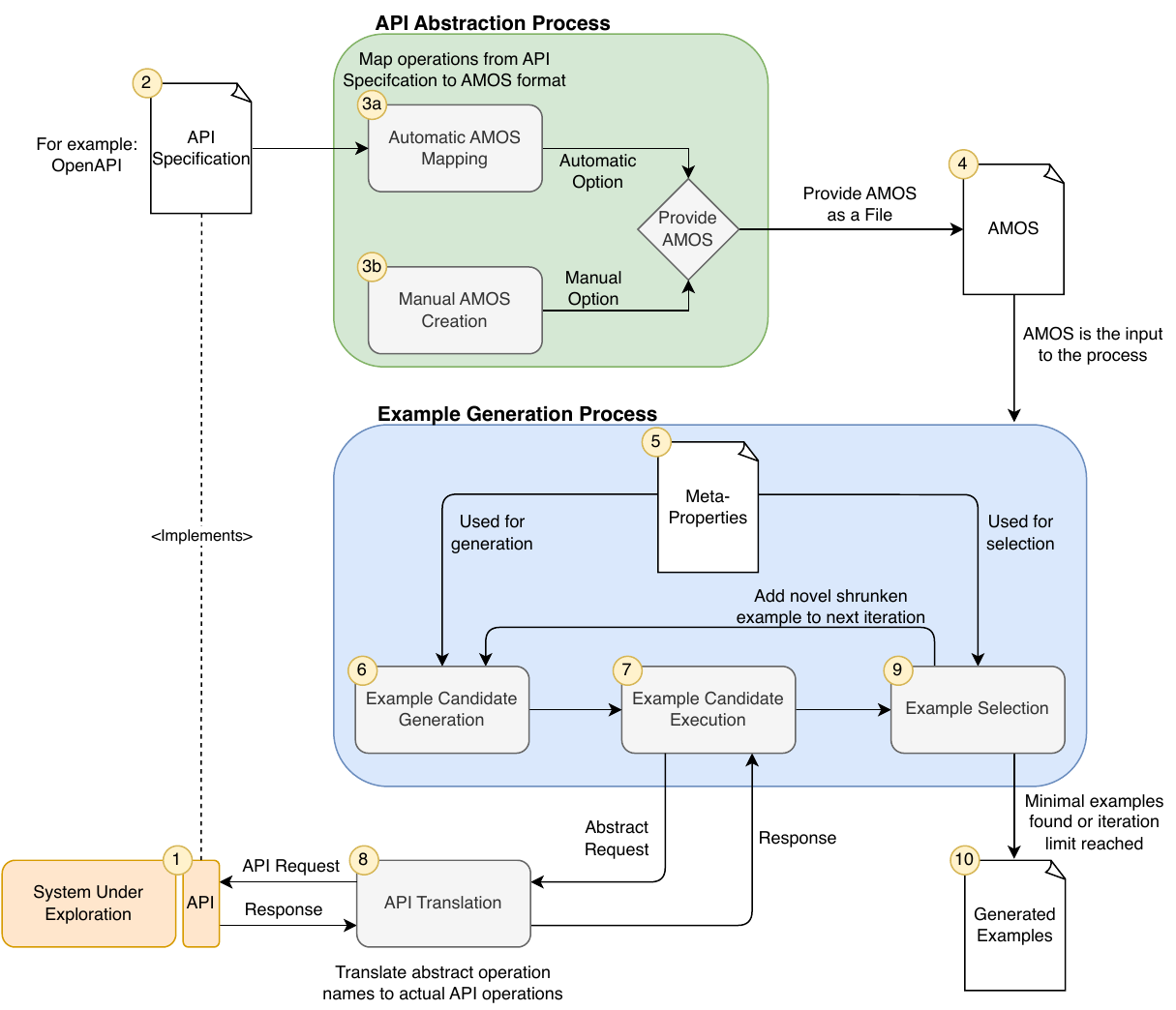}
    \caption{Approach overview; The \emph{API Abstraction} process produces an AMOS as input to the \emph{Example Generation process}. The \emph{Example Generation process} interacts with an API and produces \emph{Generated Examples}.}
    \label{fig:overview}
\end{figure*}

In this section, we expand on the key ideas and explain the details of our approach to generate relevant examples by exploring the behaviours of an API. Before describing the details of specific parts of the process, we present a schematic overview of the components and process of the proposed approach, as shown in Figure~\ref{fig:overview}.

To explore the behaviour of an API provided by a \emph{System Under Exploration}\includegraphics[scale=0.6]{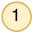}, the example generation process must be able to call the operations of the API. To be able to make any calls to an API, we need to know at least the operations it provides, and a schema of the required inputs for those operations. Without any of those, we would be forced to send random bits to the API, a process with a low probability of being useful when searching for behaviours. We aim to propose a general approach, potentially targeting any type of API implementation. Therefore, we represent the operations of the API in an abstract format, with no details of how the operations are to be executed. We refer to this format as the \textit{Abstract Method and Operations Specification} (AMOS). The details of the AMOS are further explained in Section~\ref{sec:amos}.

To automatically generate the AMOS describing the operations of an API, an \emph{API Specification}\includegraphics[scale=0.6]{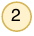} would be needed. Some examples of existing API specification formats are; OpenAPI\footnote{\url{https://www.openapis.org/}} specification (WebAPIs), a GraphQL\footnote{\url{https://graphql.org/}} schema (WebAPIs), or an Async API\footnote{\url{https://www.asyncapi.com/}} specification (message-driven APIs). This kind of specification could potentially be used to generate an AMOS automatically. However, constructing a component of \emph{Automatic AMOS Mapping}\includegraphics[scale=0.6]{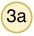} is a one-time engineering effort for the specific type of API, e.g. once a mapper from OpenAPI specifications to AMOS exists, then any REST API \cite{fielding-REST-2000} described using OpenAPI can use it. Alternatively, if no such mapper exists for a type of API, then the AMOS can be created manually\includegraphics[scale=0.6]{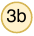}, since it is expressed in a human-readable notation. Both of these options can be seen in Figure~\ref{fig:overview} in the \emph{API Abstraction} part of the process---the goal of which is to provide an AMOS. The AMOS\includegraphics[scale=0.6]{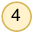} is an input to our approach and only includes the API operations, not how the operations relate, i.e., no behaviour---that is what our API exploration aims to find.

Given an AMOS as input and a set of \emph{Meta-Properties}\includegraphics[scale=0.6]{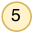}---descriptions of general API behaviours, detailed in Section~\ref{section:exploration-properties}---examples are generated by the \emph{Example Generation Process}, further explained in Section~\ref{sec:example-generation}. This process generate an example candidate\includegraphics[scale=0.6]{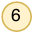}, which is then executed on the API\includegraphics[scale=0.6]{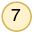}. As the \emph{Example Generation Process} only uses abstract operations---as specified in the AMOS\includegraphics[scale=0.6]{figures/callouts/4.pdf}---the abstract candidate operations must be translated to actual API operations. This translation process is done by an \emph{API Translation}\includegraphics[scale=0.6]{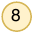} component. The response of an executed candidate operation sequence is selected\includegraphics[scale=0.6]{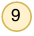} according to the conformance of the candidate to \emph{Meta-Properties}\includegraphics[scale=0.6]{figures/callouts/5.pdf}. Example candidate operation sequences which pass the selection criterion, are included in the \emph{Generated Examples}\includegraphics[scale=0.6]{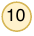}.

This overview has introduced the overall process in the proposed approach, and the components involved. In the following sections, we will describe these parts in more detail.

\begin{figure}[h]
  \centering
\begin{minted}{clj}
{:amos/id   1
 :amos/name "Demo App"
 :amos/method :clojure-api ;; <= method of invocation
 :amos/data-specs
 [{:data-spec/id   1
   :data-spec/name "Person"
   :data-spec/key  :entity/person
   :data-spec/spec [:map
                    [:person/name string?]
                    [:person/age [:and int? [:> 0]]]]}]
 :amos/operations ;; <= a list of available operations
 [{:operation/id             1
   :operation/name           "Get Persons"
   :operation/key            :get-persons}
  {:operation/id             2
   :operation/name           "Create Person"
   :operation/key            :post-person
   :operation/parameter-spec [:ref :entity/person]} ;; <= param spec
  {:operation/id             3
   :operation/name           "Delete Person"
   :operation/key            :delete-person
   :operation/parameter-spec string?}]}  ;; <= param spec
\end{minted}
    \caption{AMOS in edn format}
    \label{fig:amos}
\end{figure}

\subsection{Abstracting The System With AMOS}\label{sec:amos}

The Abstract Method and Operations Specification (AMOS)\includegraphics[scale=0.6]{figures/callouts/4.pdf} provides a general format for specifying available operations for the generation of examples. Such a specification is necessary when building a method and tool useful on many different types of SUT such as web-services, user interfaces, and libraries. One of the problems we aim to solve with this work is the burden of specifying SUT behaviour by a human. Thus, it is important to consider the effort of creating the AMOS. We must not move the burden from specifying the SUT itself to specifying the AMOS for the SUT. Since the AMOS only contains operations possible to perform on the SUT, the specification is not as detailed as a behavioural specification of a SUT, hence reducing effort. Since the AMOS does not specify behaviour, the effort of mapping the operations of the SUT to the AMOS can potentially be automated, given that the SUT conforms to a common format. The domain of web services using the OpenAPI specification is an example of where the creation of an AMOS could be automated.

Figure~\ref{fig:amos} shows the AMOS for an example application used to illustrate our approach, that will be introduced in Section~\ref{sec:running-example}. The AMOS is described in extensible data notation (edn)\footnote{\url{https://github.com/edn-format/edn}}. The essential components of the AMOS are as follows.
\begin{itemize}
    \item Method of invocation - \textit{Which API Translation component\includegraphics[scale=0.6]{figures/callouts/8.pdf} should be used to perform the executions?} Line 3 in Figure~\ref{fig:amos}.
    \item Operations - \textit{Which operations are available?} Lines 11-22 in Figure~\ref{fig:amos}.
    \item Parameter specification - \textit{What is the shape of required parameters to execute the operations?} Lines 18 and 22 in Figure~\ref{fig:amos}, where the example on Line 18 refers a specification defined on Line 5-10.
\end{itemize}

In addition to the essential components required to execute any successful operations, the AMOS can also serve as a place for further enrichment. The more we learn of a SUT, the more information can be added, either based on a human user judging generated examples or by the tool itself. An example of such additional information is whether an operation is a query operation that leaves the state of the SUT unchanged.

To specify the input parameter schema, we use a format inspired by the \emph{Malli}\footnote{\url{https://github.com/metosin/malli}} schema library. The schema allows input data structures to be defined in a system-agnostic way. In addition, this schema provides enough information to allow for the automatic creation of input generators used in the example generation process.

\subsection{Example Generation Process}\label{sec:example-generation}

The exploration of API behaviour is performed automatically based on a set of general API \emph{Meta-Properties}\includegraphics[scale=0.6]{figures/callouts/5.pdf}. The basic exploration process is the following; we start in the current state of the SUT, this state is then transformed by executing a sequence of API operations, and we observe the effects of the execution. The execution of the sequence transforms the starting state of the SUT into the final state.  

We observe changes to the system state in two different ways. The most direct way is to query the state before and after each operation; in this way, state-changes are directly observable. But being able to query the state of the system requires a query operation which tells us the current state of the system, and at the beginning of exploration, we have no such operation available. However, we can observe state-changes indirectly; when the {\em same} operation is repeated at different points in the execution sequence but returns {\em different} results, we can infer that the state of the system must have changed between the two.

The \emph{Example Generation Process} generates sequences of operations, in the \emph{Example Candidate Generation}\includegraphics[scale=0.6]{figures/callouts/6.pdf}. A Property-based testing \cite{Claessen-QuickCheck-2000} library is used for the generation of these sequences. Since our prototype was created using the Clojure programming language, we have used the QuickCheck-inspired library test.check\footnote{\url{https://github.com/clojure/test.check}}. The generated example candidate are executed on the SUT in the \emph{Example Candidate Execution}\includegraphics[scale=0.6]{figures/callouts/7.pdf}. The sequence generated depends on the specific \emph{Meta-Property}\includegraphics[scale=0.6]{figures/callouts/5.pdf} used to generate the example candidate. For example, some properties might describe the general behaviour that there should be a difference in the system state after executing a sequence of API operations. Therefore, the candidate generation process needs to generate sequences where the first and last operations are the same, so that a difference in their results can be detected by comparing their responses. 

Generated candidate examples that match a \emph{Meta-Property}\includegraphics[scale=0.6]{figures/callouts/5.pdf} in the \emph{Example Selection}\includegraphics[scale=0.6]{figures/callouts/9.pdf}, are shrunk to a minimal example by the Property-based testing library. The shrinking process re-tests smaller and smaller examples by reducing the number of operations and trying different parameters. The Property-based testing library provides sufficient shrinking algorithms for basic parameter values, but in addition we need to provide a method to shrink examples, so that the shrinking process produces examples conforming to the meta-property. The shrinking process should not remove an operation essential to the defined meta-property, nor values that other operations refer to---which would break the example. The example generation process is repeated over several iterations, where the goal of each iteration is to generate an example not previously seen. Thus, all known shrunken examples are input to the next generation iteration. In this way, the process creates a growing list of new examples previously not seen for the given \emph{Meta-Property}. The final set of selected examples are the output of the process, in the form of \emph{Generated Examples}\includegraphics[scale=0.6]{figures/callouts/10.pdf}.

The description of the \emph{Example Generation Process} shows how the effort to apply the approach to a specific type of API depends on whether previous engineering work has been done on the components for the specific target type of API---the \emph{Automatic AMOS mapping}\includegraphics[scale=0.6]{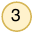} and the \emph{API Translation}\includegraphics[scale=0.6]{figures/callouts/8.pdf}. The \emph{Example Generation Process} of the approach is abstracted from the API and is thus the same, but the API-type-specific components will most likely vary in implementation effort for different types of APIs.

\subsubsection{Generating Operations}\label{sec:generating-operations}

The \emph{Example Candidate Generation}\includegraphics[scale=0.6]{figures/callouts/6.pdf} component generate calls to the API in several different ways, leveraging information about the abstract operations in the AMOS\includegraphics[scale=0.6]{figures/callouts/4.pdf}. We distinguish three different types of abstract operations, based on the way the operation's input parameters are generated. The first represents operations with randomly generated input. The second represents operations that, as their input, select a value from the parameter of a previous operation. The third represents operations that select as their input a value from a previous operation's response. 

In the exploration of the behaviour of the system, the process generates\includegraphics[scale=0.6]{figures/callouts/6.pdf} and executes\includegraphics[scale=0.6]{figures/callouts/7.pdf} a sequence of operations. Each type of abstract operation parameter generation has its own precondition. The precondition is a predicate that, given the generated execution sequence of operations so far, decides what types of operation parameters can be generated---one of the three types mentioned above. For example, when generating an operation with a parameter which is a reference to a previous operation, then the sequence so far cannot be empty.

Each operation in the generated sequence of abstract operations, selected from the AMOS, is transformed into a concrete operation for execution by the \emph{API Translation}\includegraphics[scale=0.6]{figures/callouts/8.pdf} component. Any references (to other operations' parameters or responses) can be resolved during this execution since, after execution of the operation referenced, we have the concrete value. It might sound as though creating a \emph{API Translation} component is a lot of work, but, as with the \emph{Automatic AMOS Mapping}\includegraphics[scale=0.6]{figures/callouts/3a.pdf} component, mapping specifications to AMOS, it can be done once per API type, as it is a pure translation between the abstract operation and how operations are executed in the given API type. Once a translation is created for REST APIs, for example, then it can be reused on all REST APIs, since all abstract operations would be mapped to HTTP methods. 

When the generated sequence of operations has been transformed and executed, then the responses are checked for the specific \emph{Meta-Property} of interest. This corresponds to the \emph{Example Selection}\includegraphics[scale=0.6]{figures/callouts/9.pdf} part of the process, in Figure~\ref{fig:overview}.

\subsubsection{Symbolic References}\label{section:symbolic-references}

\begin{figure*}
    \centering
    \includegraphics[width=\textwidth]{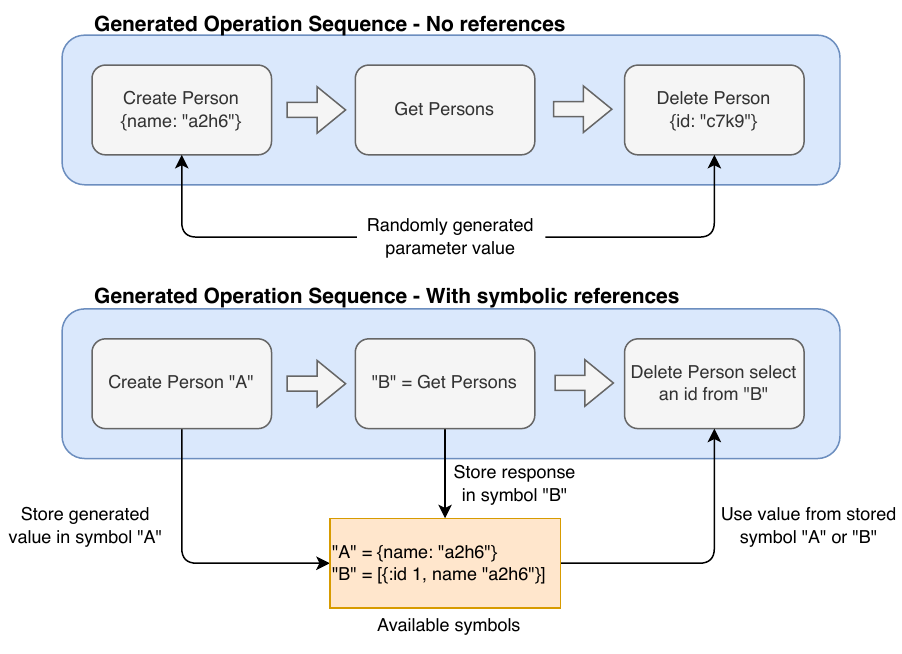}
    \caption{Two generated example sequences; one with randomly generated parameters and one which uses symbolic references for some parameters.}
    \label{fig:symbolic-refs}
\end{figure*}

Consider an API where there are dependencies between different operations. For example, it might be that to successfully delete an entity, we must use an existing identifier, which was given when the entity was created. To be able to generate such examples, we require a method of referring to previous values in the example. Such references can be to previous received responses in the example or to previous generated input parameters. As explained above in Section~\ref{sec:generating-operations}, these different types of reference are captured in different types of parameter generation for operations.

References are a means of enabling the proposed approach to generate operations with a dependency on other values---previous parameters or responses. Consider the candidate example in Figure~\ref{fig:symbolic-refs}: in this example a sequence of three operations have been generated where a ``Person'' is first created, all available ``Person'' entities are then queried, and then a deletion is performed. However, to create a ``Person'' the operation parameter must provide a ``name'' (randomly generated as ``a2h6'' in this example), but to delete a ``Person'' an ``id'' must be provided. The ``id'' of a created ``Person'' is created by the system, not the client. The only way for the client to get the ``id'' is to perform a query of the ``Person'' entities in the system. Thus, to successfully delete the created entity, the ``id'' of the created entity must be provided.

Generating such references can be done in several ways. The simplest way is to randomly generate the input parameters, which might occasionally generate the required referred value---as in the top example sequence using no references in Figure~\ref{fig:symbolic-refs}. A somewhat better way is to randomly select, as input to the current operation, any previously seen value---in Figure~\ref{fig:symbolic-refs} this would mean that the parameter of ``Delete'' would be randomly selected from ``A'' or ''B''. However, this could often select references that are not valid inputs for the current operation (as in the case of ``A'', which does not contain a required ``id'' for the ``Delete''). Finally, the method with the highest probability of selecting a relevant reference is to base the choice on the actual schema of responses and parameters (only ``B'' would match the schema for ``Delete'', since an ``id'' is required). The downside of this is that the user must provide a schema for responses (remember, we already have the schema for input in the AMOS\includegraphics[scale=0.6]{figures/callouts/4.pdf}, since those are required for any calls to the API to be meaningful). 
The default reference method that we use is to generate references that match the operation schema. Since a schema for the response of an operation is optional, we fall back to random selection of previously seen values if no schema is available.

Note that we sometimes need the concrete value of a reference, and sometimes its symbolic representation. At execution time\includegraphics[scale=0.6]{figures/callouts/7.pdf}, we need to pass the API\includegraphics[scale=0.6]{figures/callouts/8.pdf} a concrete value. However, if we execute the same example several times, then the concrete values in the system may be different each time. Thus, in order to supply the correct concrete value during re-execution, we need to retain references in their symbolic form when we report \emph{Generated Examples}\includegraphics[scale=0.6]{figures/callouts/10.pdf}. When re-executing a generated example, the symbols expressed in the example can be substituted for the actual concrete value in the specific execution.

The value of symbolic references to the approach will differ depending on the API being explored. If the responses and parameters of the API never relates, then references are not needed. However, for APIs where operations do relate, it is an essential part of the approach. As described with the example in Figure~\ref{fig:symbolic-refs}, the probability of randomly finding a successful example with the correct random values is far lower than referring to prior seen and generated values. 

\subsection{Meta-properties}\label{section:exploration-properties}

In this section, we define a set of \emph{Meta-Properties}\includegraphics[scale=0.6]{figures/callouts/5.pdf} that we use to evaluate our approach by exploring an API. As described, our approach uses meta-properties as definitions of general behaviours. 
The proposed approach does not depend on any specific kind of meta-properties, but we need some in order to evaluate the approach. We propose a set of exploration meta-properties that are general in the sense that the properties do not make assumptions about the type of the SUT (web-service, user-interface, library, etc.) or any internal (white-box) information, such as source code or usage examples.

In the most general case, observations are formulated without knowledge of what responses from the SUT mean (recall that we do not have any domain knowledge at this point). Hence, observations in the most general case are defined as the difference between the SUT state before and after the execution of an operation in the exploration example candidate sequence, or a difference in the response of an operation.

Given the two different possible ways of observing state changes, (querying the state of the system, or analysing the response of an operation) there are two categories of meta-properties. It is worth noting that when we start the exploration \textit{we do not know how to query the state of the SUT!} Hence we can not use the meta-properties defined in terms of changes to the system state until we have discovered a potential candidate query operation using meta-properties based on operation response changes. The conclusion of this reasoning is that, to be able to find system state-changing operations, we must first identify candidate query operations. 

We will now propose a set of general meta-properties. The proposed properties are based on intuitions about how Create, Delete, and Query operations should behave. We do not claim this to be an exhaustive list, but a good starting point for evaluation of the usefulness of the proposed approach. 

\subsubsection{Meta-properties based on response changes}\label{sec:meta-properties-based-on-reponse-changes}

The set of general properties based on the proposed operation response changes is as follows;

\begin{itemize}
    \item \textbf{MP-R-1: Response Equality} - \textit{Is the observed response for an execution sequence of the same operation with the same parameters equal?}
    
    We would expect a query operation to return the same response if executed multiple times in a row with the same arguments.
    \item \textbf{MP-R-2: Response Inequality} - \textit{A property of a query-like operation is that there may be a difference in the response, if the operation is executed before and after a sequence of state changing operations.}
    
    We expect this property to result in examples showing that when a create operation (for example) is called between two query operations, the responses of the query operations differ.
\end{itemize}

The main purpose of MP-R-1 and MP-R-2 is to suggest probable query operations. Recall that to formulate properties on parts of the system state, we need to know how to actually query the state.

\subsubsection{Meta-properties based on state changes}\label{sec:meta-properties-based-on-state-changes}
The proposed set of general properties based on system state changes, given a probable query operation, is the following;
\begin{itemize}
    
    \item \textbf{MP-S-1: State Identity, without observed state change} - \textit{Is there a sequence of operations that, given state S, after execution results in the same state S and \textbf{no observed} state changes took place?}
    
    An example generated by this property would only contain non-state-changing operations in the execution sequence, which might indicate query-like operations, or operations that sometimes cause state changes but failed. The main difference between this property and MP-R-1 is that in MP-R-1 the sequence consists of the same operation with the same parameters, while in this property the sequence can contain any operation with any parameters. \\
    
    \item \textbf{MP-S-2: State Mutation} - \textit{Is there a sequence of operations that, given state S, after execution results in a state not equal to S?}
    
    Operation sequences that match this property would include state-changing operations, which could be creations, deletions, or both.\\
    
    \item \textbf{MP-S-3: State Identity, with observed state change} - \textit{Is there a sequence of operations that, given state S, after execution, results in the same state S, but we observed a change in state during execution?}
    
    An example generated by this property would contain at least one state-changing operation in the execution sequence. In addition, if we start from an empty state, the sequence would contain both a create operation, resulting in an observed state change, and a delete operation, bringing the state back to the identity state.
    
\end{itemize}
    
\subsubsection{Meta-properties based on state change and size}

The final meta-properties are based not only on state-changes, but also on the \textit{size} of the state-change. To measure the size in a SUT-agnostic way, we use the length of the compressed (gzip) response. We compress the response to reduce the weight of the common structure and increase the weight of the novel values.
    
    \begin{itemize}
    \item \textbf{MP-S-4: State Mutation, State Increase} - \textit{Is there a sequence of operations that, given state S, after execution, results in a state not equal to S, and the size of the state has increased?}
    
    Examples generated by this property will contain at least one operation increasing the state, i.e., a create-like operation. \\
    
    \item \textbf{MP-S-5: State Mutation, State Decrease} - \textit{Is there a sequence of operations that, given state S, after execution results in a state not equal to S, and the size of the state has decreased?}
    
    This property will generate examples where the execution sequence contains at least one operation reducing the state, i.e., delete-like. For this property to make sense, at least when the starting state is empty, we must first generate and execute at least one operation before the rest of the execution sequence. It is not possible to successfully delete an entity if it has not been created first. All such operations are included in generated examples of this property since they are relevant to the example. \\
    
\end{itemize}

\begin{figure}[h]
  \centering
\begin{minted}{clj}
;; Example in raw edn-format
[{:operation/key
  {:operation.key/value  :post-person
   :operation.key/symbol 'a}
  :operation/parameter
  {:operation.parameter/value  {:person/name "0" :person/age 65}
   :operation.parameter/symbol 'b}}
 {:operation/key
  {:operation.key/value  :get-persons
   :operation.key/symbol 'c}
  :operation/parameter  {:operation.parameter/symbol 'd}}
 {:operation/key
  {:operation.key/value  :delete-person
   :operation.key/symbol 'e}
  :operation/parameter
  {:operation.parameter/value            "0"
   :operation.parameter/path             [0 :person/name]
   :operation.parameter/symbol           'f
   :operation.parameter/reference-symbol 'c}}]

;; Same example in 'humanized' form
["State S = response from the given query operation"
 "execute :post-person with #:person{:name "0", :age 65}"
 "a = response from executing :get-persons"
 "execute :delete-person with 0, selected from [0 :person/name] in a"
 "when the query operation is executed again, the response is 
  equal to S"]

;; Same example as a synthesized Clojure test
(deftest state-identity-with-state-change
  (let [_      (execute-fn :reset)
        s-pre  (execute-fn :get-persons)
        _      (execute-fn :post-person #:person{:name "0", :age 65})
        a      (execute-fn :get-persons)
        _      (execute-fn :delete-person 
                 (get-in a [0 :person/name]))
        s-post (execute-fn :get-persons)]
    (is (= s-pre s-post))))
\end{minted}
    \caption{Reporting}
    \label{fig:reporting}
\end{figure}

\subsection{Reporting Examples}\label{section:reporting-examples}

The result of the example generation process is that after the given iterations, several examples are produced\includegraphics[scale=0.6]{figures/callouts/10.pdf} (unless no new novel examples were found). As we are aiming to increase understanding of the software, it is important not to overwhelm the user with too many examples. We prefer shorter examples to longer ones because we want to show an example of the meta-property with as high a signal-to-noise ratio as possible. We reason that if a shorter sequence can produce a specific behaviour, it is preferable to a longer sequence with the same property.

The reporting and presentation of the examples are important if they are to be of use to a human user. In addition, it is also beneficial for the output to be machine-readable, for any future report extensions. How humans best want to be presented with examples needs further empirical investigation and is beyond the scope of this paper. However, to present the examples in the evaluation of this paper, we have built a proof-of-concept reporter producing a report in a human-readable representation.

Figure~\ref{fig:reporting} shows three different ways of reporting the same generated example. First, the data structure that represents it (L2-19), in human-readable edn-format (the same format as used in the AMOS). Second, the example is formatted as a list of strings, which is more accessible to a human reader---using our ``humanised'' reporter. Note the use of a symbolic reference (as explained in Section~\ref{section:symbolic-references}) in L24-25. The use of the symbol in L25 shows both the concrete value for this execution and the path in the value referenced by the symbol. As an example of how a \emph{Generated Example} can be transformed for multiple uses, the final output is in the form of a test case capturing the behaviour of the meta-property used to produce this example. In this case, the usage of symbolic references is essential. If we relied on the concrete value, we might not be able to rerun the test case, depending on the system state.


\section{An Illustrative Exploration}\label{section:an-exploration-example}

In this section, we apply the proposed approach to a small API to illustrate how the approach works, before going into our evaluation. We limit the exploration to the meta-properties based on responses (MP-R-1 and MP-R-2 from Section~\ref{sec:meta-properties-based-on-reponse-changes}) and save the use of the other properties for the evaluation in Section~\ref{section:evaluation}.

\subsection{The Application}\label{sec:running-example}
To illustrate our approach, we use a stateful application as an example. The example represents a simple API implementation including the basic operations for managing entities: post, get, and delete. The application provides the following API operations:

\begin{itemize}
    \item \texttt{post-person} - Adds the provided person to the database, indexed by their name. The initial implementation does not check whether the person to be added already exists in the database or not---it will just be overwritten if it already exists. This lack of input validation is a behaviour for which our method can produce an example and is something we will address when we explore the behaviour of this application in Section~\ref{sec:exploration-of-approach}.
    \item \texttt{get-persons} - Return the persons in the database or an empty list if the database is empty.
    \item \texttt{delete-person} - Deletes a person with the given name. The initial implementation does not check whether the person to be deleted exists in the database or not.
\end{itemize}

For the interested reader, a walk-through of the source code of the running example is provided in Appendix~\ref{app:code-walkthrough}. However, understanding the code is not essential to understand how the proposed approached is applied.

\subsection{The Exploration}\label{sec:exploration-of-approach}

We run through an exploration by interactively applying our proposed approach, as shown in Figure~\ref{fig:interactive-process}. Starting from the application described in Section~\ref{sec:running-example}, we iteratively update the application based on the generated examples and the realisations they enable.

When generating the examples in this exploration, the state of the system is reset between generations. A reset operation key is sent to the \emph{API Translation} component, as with any other operation, and it is up to the translator to perform the reset of the SUT if possible. Our approach can be used with and without an implemented reset function. However, to keep this exploration simple, we use a reset function and discuss the differences caused by not having one in the evaluation in Section~\ref{section:evaluation}.

As we described in Section~\ref{section:exploration-properties}, the meta-properties based on system state changes require a query operation. However, at the start of the exploration, we do not yet know which operations show a Query-like behaviour. Therefore, the first step in exploration is to find candidate query operations, but in that process we might also discover other behaviours. We do so by generating examples conforming to ``Response equality'' (MP-R-1) and "Response inequality" (MP-R-2).

To put the exploration into the context of the approach overview in Figure~\ref{fig:overview}; the input to the \emph{Example Generation Process} is the AMOS\includegraphics[scale=0.6]{figures/callouts/4.pdf} describing the operations of the API we explore. In this particular case, the AMOS used is shown in Figure~\ref{fig:amos}. However, its details are not important for the purpose of understanding the exploration results. In addition, the user selects the meta-properties to explore, in this case, as mentioned above, we start with MP-R-1 and MP-R-2.
Figure~\ref{fig:exploration-example-1} shows the result. For readability, the result is shown in our "humanised" string format. Transforming the output data-format of the generated examples into a richer presentation format, such as PDF, HTML or similar, is just an engineering effort and not central to understanding the proposed method.

\begin{figure}[h]
  \centering
\begin{minted}{clj}
;; Response inequality
["Executing :get-persons"
 "then executing :post-person with #:person{:name "", :age 1}"
 "and then executing :get-persons again, results in different 
  responses"]

;; Response equality
["Executing :post-person with #:person{:name "", :age 1} twice 
  results in equal responses"]

["Executing :delete-person with an empty string twice 
  results in equal responses"]

["Executing :get-persons twice results in equal responses"]
\end{minted}
    \caption{First exploration result of the example application.}
    \label{fig:exploration-example-1}
\end{figure}

Looking at the result, we can observe that the \texttt{get-persons} operation conforms to the "response-inequality" meta-property (L1-4), it was the only operation doing so, and we can see the shortest shrunken example producing this behaviour. The shrinking process not only shrinks the sequence, it also tries to shrink the parameter values. This is the reason why all example names are empty strings, which is the smallest passing value. The example is very intuitive of how a query operation should behave; the response of the operation is different when an entity has been created in between two executions. This example also gives a strong indication that the \texttt{post-person} operation is a state-changing operation. Remember when looking at the examples, names of operations have no meaning to the generation method, we draw no conclusions from them---they could be any random string (but obviously, this would be much harder for humans to understand).

All three operations are found in examples of the ``response equality'' property. Why is that? The most intuitive example might be for the get-operation. We expect a query operation to give the same response if executed twice in a row with the same parameters. As for the delete- and post-operations, it might not be as clear why these examples are generated. Looking back at the description of the application we explore, in Section~\ref{sec:running-example}, recall that in the post- and delete-operations, it is not checked whether the person is actually in the database. In consequence, the delete-operation tries to remove persons whether or not they exist, with the same response to the client for the same input (which is a reasonable API behaviour), and the post-operation successfully adds the same person twice without any difference in the client result. Is this correct behaviour? We do not attempt to judge this; only the requirements of the application can state whether it is acceptable for the post-operation to add persons with the same name, potentially overwriting a person, or whether this is a bug.

Suppose that we consider the behaviour of the post-operation in this example to be a bug. We can easily fix it by adding a check before adding a person to the database: if the person exists, we return a failure instead. Let us see how this changes the exploration result.

\begin{figure}[h]
  \centering
\begin{minted}{clj}
;; Response inequality
["Executing :post-person with #:person{:name "", :age 1} twice 
  results in different responses"]

["Executing :get-persons"
 "then executing :post-person with #:person{:name "", :age 1}"
 "and then executing :get-persons again, result in different 
  responses"]

;; Response equality
["Executing :delete-person with an empty string twice results in 
  equal responses"]

["Executing :get-persons twice results in equal responses"]
\end{minted}
    \caption{Second exploration result of the example application.}
    \label{fig:exploration-example-2}
\end{figure}

The second exploration produces the result in Figure \ref{fig:exploration-example-2}. What can we learn from this? As intended, two executions of the post-operation with the same parameters now display a response inequality. The first call adds the person, while the second call fails. Also, note that we can no longer find any example of the post-operation for response equality.

This could have been the end of our exploration, but requirements for software constantly change. Suppose that our application should now only allow senior citizens to be added to the database and also that empty names are no longer considered valid. We add the following input validation predicate to the post-operation:

\begin{figure}[h]
  \centering
\begin{minted}{clj}
;; Response inequality
["Executing :post-person with #:person{:name "0", :age 65} twice 
  results in different responses"]

["Executing :get-persons"
 "then executing :post-person with #:person{:name "0", :age 65}"
 "and then executing :get-persons again, result in different 
  responses"]

;; Response equality
["Executing :delete-person with an empty string twice results in 
  equal responses"]

["Executing :get-persons twice results in equal responses"]

["Executing :post-person with #:person{:name "", :age 1} twice 
  results in equal responses"]
\end{minted}
    \caption{Third exploration result of the example application.}
    \label{fig:exploration-example-3}
\end{figure}

\begin{verbatim}
 (and (>    (:person/age  person) 64) 
      (not= (:person/name person) ""))
\end{verbatim}

Once again, we want to explore the consequences of the actual behaviour of this change. The third and final result for this exploration example is shown in Figure~\ref{fig:exploration-example-3}.

We can now observe that the post-operation has generated examples for \emph{both} the properties. The inequality property example now shows the behaviour of two consecutive calls with valid input, and the response equality example shows two invalid calls. Note that, due to shrinking, the examples of a post-operation with valid input produce the smallest input passing the boundary condition. We are now satisfied with the behaviour of our application and have concluded the exploration. 

It is worth recalling that the generation of these examples did not require any white-box information, such as source code, usage examples, or any formal specification. In summary, with this illustrative exploration, we have shown the flow of working with an example generating method, and the value of examples in conveying an understanding of how an application actually behaves and evolves through application changes.  

\section{Performance of Symbolic References}\label{section:evaluation-symbolic-refs}

When describing our approach, we have introduced the concept of \emph{Symbolic References} (Section \ref{section:symbolic-references}). When a value is generated or received as a response from the SUT, we can store these values with a symbolic name. When further operations are generated in a trial example sequence, stored symbols can be referred to, and thus previously seen values can be reused. If an example is successfully generated with the use of symbols, these can be used in the resulting example output, making it more succinct compared to repeating values.

Using this method is in contrast to using random values for all input values of the operation. A reasonable assumption is that references to values already in the sequence, or otherwise observed, should outperform random values. But is that true? Using symbolic references increases the cost of the implementation, as it is easier to rely on random values, rather than implementing a mechanism for reference storage and resolutions. Thus, there should be a significant performance improvement in the number of test cases needed in the search for an example to justify the extra implementation effort. In this section, we describe our evaluation of using references over random values in the generated sequences. We perform this experiment in order to make an informed decision on whether to use symbolic references or not in our main evaluation of an industry-grade API in Section~\ref{section:evaluation}.

\subsection{Experiment Setup}

We used the application previously described in Section~\ref{sec:running-example}, as our controlled application. We also produced a version with the incorporated input validation from Section~\ref{section:an-exploration-example}, which allows persons to be created only if their age is above 64 and their name is non-empty. 

The setup results in four different configurations denoted A-D. 
\begin{enumerate}[A --]
    \item Random parameter generation, no input validation.
    \item Random reference generation, no input validation.
    \item Random parameter generation, input validation.
    \item Random reference generation, input validation.
\end{enumerate}
``Random parameter generation'' (A and C) means that all input values of generated operations are generated as random values. ``Random reference generation'' (B and D) means that input values can refer to previously stored symbols, but the symbol to use is selected randomly.

We collected data by running each configuration 1000 times with a maximum budget of 1000 test cases to find an example. In order for references to previous values to be relevant, the meta-property used should be a state-based one. For this experiment we used the meta-property MP-S-3, ``State identity, with observed state change''. As described in Section~\ref{sec:meta-properties-based-on-state-changes}, successfully generated examples of this meta-property would contain both a create operation and a delete operation. Thus, the delete operation must be performed on the entity previously created in the sequence---making the use of symbolic references potentially useful. Data were analysed using a pairwise comparison with a \textit{Wilcoxon-Mann-Whitney} test and \textit{Varga-Delaney A measure} to measure the effect size. This setup was chosen considering the guidelines for evaluations of random-based algorithms \cite{arcuri-hitchhiker-2014}.

\subsection{Result}

\begin{figure*}
    \centering
    \includegraphics[width=\textwidth]{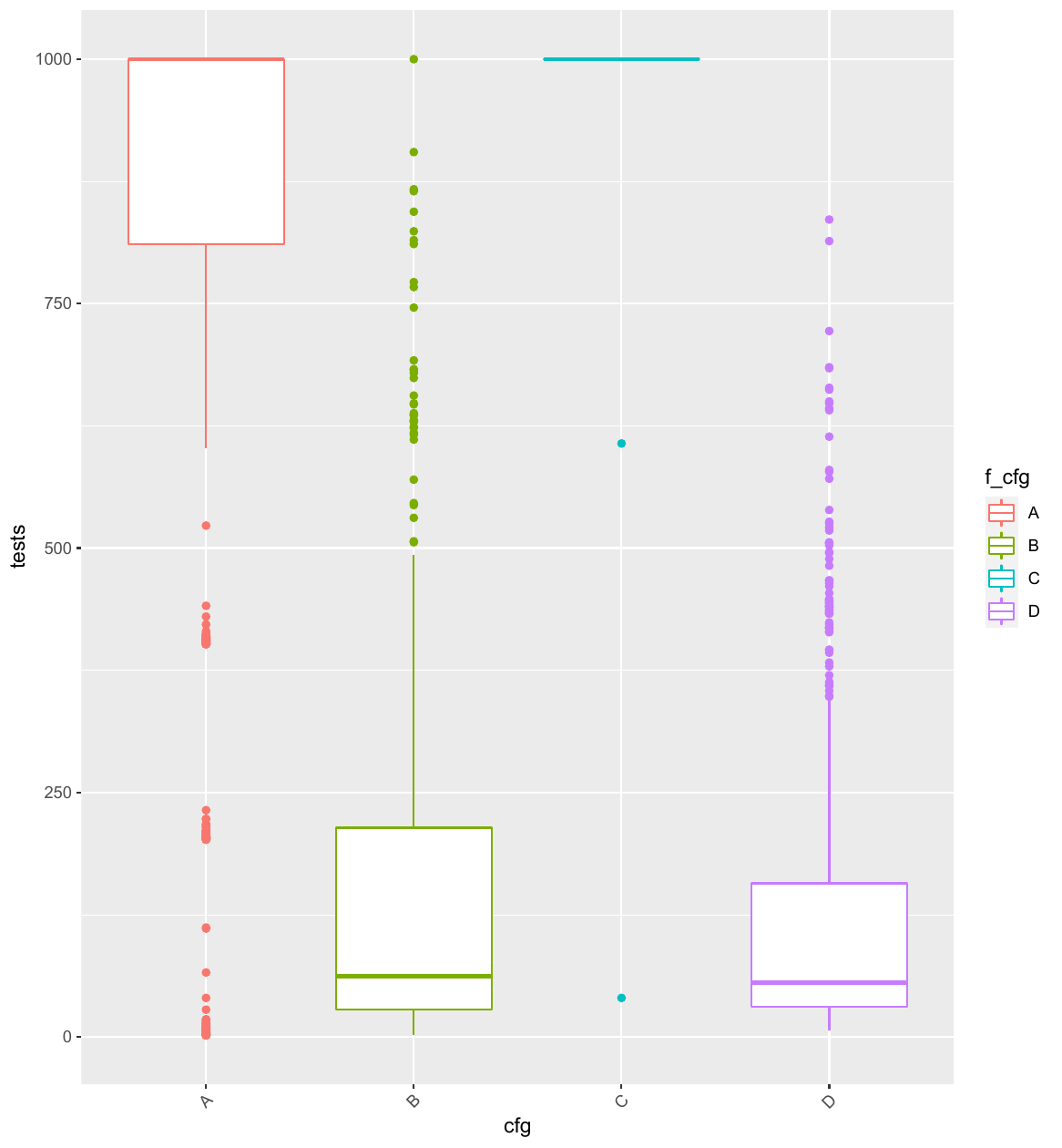}
    \caption{Box plot of the number of test cases needed for each configuration}
    \label{fig:boxplot}
\end{figure*}

\begin{table}[h]
\begin{center}
\caption{Summary statistics of test case data}\label{tab:summary}%
\begin{tabular}{@{}lllllll@{}}
\toprule
Cfg & Min & 1st Qu. & Median & Mean & 3rd Qu. & Max \\
\midrule
A & 2.0 & 811.0 & 1000.0 & 839.8 & 1000.0 & 1000.0 \\
B & 2.0 & 28.0 & 62.0 & 136.5 & 214.0 & 1000.0 \\
C & 40.0 & 1000.0 & 1000.0 & 998.6 & 1000.0 & 1000.0 \\
D & 7.0 & 31.0 & 55.5 & 116.5 & 157.2 & 836.0 \\
\bottomrule
\end{tabular}
\end{center}
\end{table}

\begin{table}[h]
\begin{center}
\caption{Wilcoxon-Mann-Whitney test and Varga-Delaney effect size}\label{tab:Wilcoxon-Mann-Whitney}%
\begin{tabular}{@{}llll@{}}
\toprule
Cfg & \textit{p}-value & $\hat{A}_{12}$ \\
\midrule
A-B & $< 2e^{-16}$ & 0.92501 \\
B-C & $< 2e^{-16}$ & 0.00117 \\
A-C &  $< 2e^{-16}$ & 0.36801 \\ 
B-D & 0.5593 & 0.50754  \\
A-D &  $< 2e^{-16}$ & 0.92925 \\
C-D &  $< 2e^{-16}$ & 0.99935 \\
\bottomrule
\end{tabular}
\end{center}
\end{table}

The statistics of the collected data are presented in Table~\ref{tab:summary}. Figure~\ref{fig:boxplot} shows the data in a graphical representation, in the form of a box plot. When visualising the data, as in Figure~\ref{fig:boxplot}, we can see a strong indication that configurations B and D (both using random reference generation) performed much better than A and C (both using random parameter generation). The visualisation also indicates that there is a big difference in the configurations using random parameter generation when input was validated, A vs. C. However, it is harder to visually see if there is a significant difference in input validation being used or not with configurations using random reference generation, B vs. D. 

Table~\ref{tab:Wilcoxon-Mann-Whitney} shows the result of the pairwise Wilcoxon-Mann-Whitney test and the effect size. The Varga-Delaney $\hat{A}$ measure is 0.5 if there is no effect of the pairs---they perform best in an equal amount of cases, 0.0 would mean that the first alternative is always better, while 1.0 that the second alternative is always better. We can see significant differences in all configurations except B-D (p-value 0.5593). In addition, there are very strong effect sizes towards B and D when compared to A and C.

The statistical analysis confirms that configurations using random reference generation are significantly better than random parameter generation, with very large effect sizes. In addition, there is also a significant difference between A and C, but not between B and D. This means that input validation has a significant effect on configurations using random parameter generation, but \textbf{not} on configurations using random reference generation.

\subsection{Discussion}
Although the application we used in this experiment is very simple and not industry grade, we see no reason that random input generation would perform better given a more complex application. Indeed, if random input generation cannot perform better in this simple scenario, we consider it unlikely that it would outperform references in an industry-grade scenario.

We can put this result in the context of our main evaluation of an industry-grade REST API (Section \ref{section:evaluation}). REST API test generation is time consuming as it is performed at the system level \cite{Zhang-Fuzzing-Microservices-In-Industry-2022}. Additionally, finding relations between operations in sequences in the generation of REST API tests is reported as one of the main challenges \cite{Kim-Automated-Test-Generation-for-REST-APIs-No-Time-to-Rest-Yet-2022}. Our results show that the number of tests required to reach a conclusion can be significantly reduced by referring to previously used and observed values in a sequence, which provides further evidence that research on how to effectively find references between operations for REST APIs is worth pursuing.

In summary, random reference generation, using stored symbolic references, requires significantly fewer tests to reach a conclusion than configurations based on parameter generation. In addition, reference generation configurations are robust to whether input validation is used or not. Our analysis gives us strong support for going forward with a random reference generation configuration if no explicit references are provided.

\section{Evaluation}\label{section:evaluation}

In this section, we will make a deeper exploration of a set of API operations of an industry-grade system. Our method is one of interactive exploration, which means that the user plays an essential role. Thus, it is not practical to show evaluations on many different systems. Instead, we show more details of performing the exploration in one case. The DevOps platform GitLab\footnote{\url{https://about.gitlab.com/}} is the selected system. As a DevOps platform, GitLab provide APIs to, for example, manage entities such as users, groups, code issues, code change requests, etc. GitLab has been the system evaluated in several case studies of REST API fault-finding methods, because it is industry-grade and available to run locally \cite{Atlidakis-RESTler-2019, Karlsson-QuickREST-2020, Wu-Combinatorial-Testing-of-RESTful-APIs-2022}. GitLab's REST API represents a RESTful API, where entities are manipulated by HTTP verbs.

The method proposed in this paper is a way to explore the behaviour of an API. Therefore, our evaluation will be an exploration of a part of the GitLab API, showing what insights the method provides about an industry-grade system.

The main research question we seek to answer is; which of the proposed meta-properties generate examples that further our understanding of the API?

\subsection{Setup}

We ran GitLab on a local installation on developer hardware.\footnote{MacBook Pro, 2.9 GHz Intel Core i9, 16 GB RAM} An AMOS was automatically created based on a prototype mapper from OpenAPI specifications, and then manually complemented where the prototype was lacking. It was created for the Group API operations, GET, POST, and DELETE. We used an \emph{API Translation} component specific to the REST API-type of API. Each meta-property was executed with 100 example tests per iteration and 5 iterations. The research prototype implementation used can be downloaded as a replication package.\footnote{\url{https://figshare.com/s/123e5bee7ec2ea893abb}}

\subsection{Exploration Result}

\subsubsection{Finding a Query Candidate}

To be able to use any of the meta-properties based on state-change, we first need a query-operation candidate. As in our running example, we use the response-based meta-properties, MP-R-1 and MP-R-2, to try to find a good candidate.

\begin{figure}[h]
  \centering
\begin{minted}{clj}
;; Response inequality
["Executing :post-groups with 
  {"name" "2U", "path" "00", "description" "64k9", 
   "membership_lock" false, "visibility" "private", 
   "share_with_group_lock" true} 
  twice results in different responses"]

["Executing :get-groups with {"page" 15}"
 "then executing :post-groups with 
  {"name" "1Ujt", "path" "10", "description" "NI9l34n", 
   "membership_lock" true, "visibility" "internal"}"
 "then executing :post-groups with 
  {"name" "Eo", "path" "", "description" "9GD" 
   "visibility" "public", "share_with_group_lock" false}"
 "then executing :delete-groups with {"id" 0}"
 "and then executing :get-groups again, result in different 
  responses"]

;; Response equality
["Executing :delete-groups with {"id" 0} twice results in 
  equal responses"]

["Executing :post-groups with {"name" "", "path" ""} twice results 
  in equal responses"]

["Executing :get-groups with {} twice results in equal responses"]
\end{minted}
    \caption{First exploration result of GitLab with no reset.}
    \label{fig:gitlab-MPR-no-reset}
\end{figure}

Figure~\ref{fig:gitlab-MPR-no-reset} shows the first exploration result. The response-inequality property shows two operations, \texttt{:post-groups} and \texttt{:get-groups}. Since there is an inequality in the response with the same input and no operation is required between them, the result for \texttt{:post-groups} indicates that there is a condition that prevents an entity from being recreated with the same parameters. This is further supported by the fact that \texttt{:post-groups} is found to have response equality for calls with empty strings for \texttt{name} and \texttt{path}.

The other operation with examples of response inequality is \texttt{:get-groups}. This example (L8-17) shows a sequence of different operations, resulting in the response of the first and second \texttt{:get-groups} being different. A query operation should not change the state by itself. To be a candidate for a query operation, there should be at least one state-changing operation between the first and second calls to the query operation. Furthermore, a query operation should have a response equality without any state-changing operation in between, which is the case for \texttt{:get-groups} (L26).

With these first exploration results, we have examples that indicate that \texttt{:get-groups} behaves as a query operation should do (response inequality with state-changing operations in between and response equality without). The \texttt{:post-groups} operation shows examples of being a state-changing operation. It is found in an example as the middle operation between (what we think are) two query operations and hence must have been the cause of the state-change. In addition, there are examples of response changes with two subsequent calls of this operation, indicating that it must be a type of operation that can change the state.

Recall that our approach aims to produce the smallest example of a behaviour. This objective is not achieved in Figure~\ref{fig:gitlab-MPR-no-reset}. For example, is the \texttt{membership\_lock} parameter really needed for \texttt{:post-groups}? The GitLab documentation\footnote{\url{https://docs.gitlab.com/ee/api/groups.html\#new-group}} states that it is not. Also, the sequence in the example on lines 8-17, showing \texttt{:get-groups} to have response inequality, contains two \texttt{:post-groups} operations and one \texttt{:delete-groups}. This is clearly not the shortest possible sequence; it should suffice with a get-post-get sequence. Why is shrinking not performing as well as we would expect? The answer lies in the difference between using a soft-reset, and not.

\subsubsection{Soft-reset}

The process of shrinking examples works best when responses from the system are deterministic. When an example is produced, in the test generation process, the shrinking process will repeat tests with simplified input. This process continues until the minimal input producing the failing test (in our case, an example still behaving according to the property we explore) is found. When behaviour is non-deterministic, the shrinking process may get different results when trying to minimise---a test that failed may now succeed, resulting in a larger than possible shrunk example. In particular, this can happen when state changes induced by one test affect the behaviour of the next.

One way to alleviate this is to use a soft-reset function. As we propose a black-box method, we do not want the reset function to depend on internal knowledge of the application. The implementation of the reset function is injected---thus, external to the core exploration process of our proposed approach---and system dependent, but it can still be implemented in a black-box manner using the external API of the system. To the internals of our approach, this is just an abstract ``reset'' operation passed to the translation components between the core process of our approach and the SUT. This is what we did in the GitLab case. By querying all the available entities for a specific type, groups in our case, we can then call delete on all groups received. Thus, we do not depend on any white-box information on how GitLab is implemented, although we do need to know how to delete groups.

Figure~\ref{fig:gitlab-MPR-reset-no-sleep} shows the updated result when exploring with a soft-reset function.

\begin{figure}[h]
  \centering
\begin{minted}{clj}
;; Response inequality
["Executing :get-groups with {"page" "2"} twice results in different 
  responses"]

["Executing :post-groups with {"name" "0", "path" "00"} twice results 
  in different responses"]

;; Response equality
["Executing :delete-groups with {"id" 0} twice results in 
  equal responses"]

["Executing :post-groups with {"name" "", "path" ""} twice results 
  in equal responses"]

["Executing :get-groups with {} twice results in equal responses"]
\end{minted}
    \caption{Exploration result of GitLab with reset.}
    \label{fig:gitlab-MPR-reset-no-sleep}
\end{figure}

The result is much shorter, but does it tell the same story? The examples related to \texttt{:post-groups} (L5-6, L12-13) look similar to those without the reset function, but smaller. However, there is a difference in the result for the \texttt{:get-groups} operation. It now has response-inequality with only two calls, which is not expected of a candidate query operation. After some investigation, we discovered that the deletion operation for groups is asynchronous. This means that when groups are deleted in the soft-reset, there is not enough time for them to be fully deleted before the next test starts. The result is that the state is still changing between calls to \texttt{:get-groups}, and so the response is different and the example is stored. This illustrates the problems that asynchronous operations can cause for automated API exploration. We fix this by adding a short sleep at the end of our black-box soft-reset, to allow the deletions to complete.

\begin{figure}[h]
  \centering
\begin{minted}{clj}
;; Response inequality
["Executing :get-groups with {}"
 "then executing :post-groups with {"name" "0", "path" "00"}"
 "and then executing :get-groups again, result in different 
  responses"]

["Executing :post-groups with {"name" "0", "path" "00"} twice results 
  in different responses"]

;; Response equality
["Executing :delete-groups with {"id" 0} twice results in equal 
  responses"]

["Executing :post-groups with {"name" "", "path" ""} twice results in 
  equal responses"]

["Executing :get-groups with {} twice results in equal responses"]
\end{minted}
    \caption{Exploration result of GitLab with reset and sleep.}
    \label{fig:gitlab-MPR-reset-with-sleep}
\end{figure}

Figure~\ref{fig:gitlab-MPR-reset-with-sleep} shows the result. These examples are the sharpest so far and also minimal. These three explorations show the trade-off between using a soft-reset or not. The method works without one, but tests run more slowly because the state grows larger, and the resulting examples may not be minimal.

\subsubsection{State-change meta-properties}

For the state-changing meta-properties, we need to provide a query operation. Based on the results of MP-R-1 and MP-R-2, discussed in the previous sections, the \texttt{:get-groups} operation is the one with examples expected of a query operation.

MP-S-1 (State Identity, without observed state change),  with results in Figure~\ref{fig:gitlab-MPS1}, shows another example expected of the get and post-operation, the get-post-get sequence where the parameters of the post-operation are outside the previously discovered input boundary (empty name and path). The other identity sequences show examples getting and an unsuccessful delete, also resulting in no state change. In summary, the new information we obtained from the exploration is that unsuccessful post and delete operations do \textit{not} change the state.

\begin{figure}[h]
  \centering
\begin{minted}{clj}
;; state identity without state change
["State S = response from the given query operation"
 "execute :delete-groups with {"id" 0}"
 "when the query operation is executed again, the response is equal 
  to S"]

["State S = response from the given query operation"
 "execute :post-groups with {"name" "", "path" ""}"
 "when the query operation is executed again, the response is equal 
  to S"]

["State S = response from the given query operation"
 "execute :get-groups with {}"
 "when the query operation is executed again, the response is equal 
  to S"]
\end{minted}
    \caption{Exploration result of GitLab with MP-S-1.}
    \label{fig:gitlab-MPS1}
\end{figure}

The State Mutation (MP-S-2) property in Figure~\ref{fig:gitlab-MPS2}, shows two executions. The first execution shows the result with our best candidate for a query operation, \texttt{:get-groups}, the other shows the result if we provide \texttt{:post-groups} as another query candidate---which a user exploring the system might do. With \texttt{:get-groups} as a candidate, we get the expected get-post-get examples (with some non-state-changing operations in some of the sequences), but this was already indicated by MP-R-1, so we get no new information. What may be more disturbing is the result of using \texttt{:post-groups} as the query candidate. Lines 25-28 show an example where a first successful post is executed, followed by a non-successful post, and then another post with the same parameters as the first. The second call will fail since, as we have discovered previously, we cannot post the same parameters twice. As this property seeks for changes between the first and last operations with a sequence in between, this example matches that pattern. However, this property is not doing a good job of providing relevant new examples.

\begin{figure}[h]
  \centering
\begin{minted}{clj}
;; state mutation with :get-groups as query operation
["State S = response from the given query operation"
 "execute :delete-groups with {"id" 0}"
 "execute :post-groups with {"name" "0", "path" "00"}"
 "when the query op. is executed again, the response is NOT equal 
  to S"]

["State S = response from the given query operation"
 "execute :get-groups with {}"
 "execute :post-groups with {"name" "0", "path" "00"}"
 "when the query op. is executed again, the response is NOT equal 
  to S"]

["State S = response from the given query operation"
 "execute :post-groups with {"name" "0", "path" "00"}"
 "when the query op. is executed again, the response is NOT equal 
  to S"]

;; state mutation with :post-groups as query operation
["State S = response from the given query operation"
 "execute :delete-groups with {"id" 0}"
 "when the query op. is executed again, the response is NOT equal 
  to S"]

["State S = response from the given query operation"
 "execute :post-groups with {"name" "", "path" ""}"
 "when the query op. is executed again, the response is NOT equal 
  to S"]

["State S = response from the given query operation"
 "execute :get-groups with {}"
 "when the query op. is executed again, the response is NOT equal 
  to S"]
\end{minted}
    \caption{Exploration result of GitLab with MP-S-2.}
    \label{fig:gitlab-MPS2}
\end{figure}

\begin{figure}[h]
  \centering
\begin{minted}{clj}
;; state identity with state change
["State S = response from the given query operation"
 "execute :post-groups with {"name" "0", "path" "00"}"
 "a = response from executing :get-groups"
 "execute :delete-groups with {"id" 41600}, selected from [0 :body 0] 
  in a"
 "when the query operation is executed again, the response is equal 
  to S"]
\end{minted}
    \caption{Exploration result of GitLab with MP-S-3.}
    \label{fig:gitlab-MPS3}
\end{figure}
Executing MP-S-3 (State Identity, with observed state change) in Figure~\ref{fig:gitlab-MPS3}, results in our first example with a response reference. To be able to successfully delete an entity and get back to the identity state, the delete operation must be provided with an existing id. In GitLab, these ids are created on the server side. As can be seen in this generated example, an entity is created with the \texttt{post-operation}, then \texttt{get-groups} is called with the response stored in the symbol \texttt{a}. Finally, \texttt{delete-groups} is called, with a value selected from the previous response and stored in the symbol \texttt{a}. This sequence has a very high signal-to-noise ratio since it shows the interaction of all three operations.

The last two properties, MP-S-4 and MP-S-5, are also based on a state mutation like MP-S-2, i.e., the state differs between the first and last operations. However, MP-S-4 and MP-S-5 add the additional check that the state should not only be different, but also differ in size.

\begin{figure}[h]
  \centering
\begin{minted}{clj}
;; state mutation with increase, :get-groups as query operation
["State S = response from the given query operation"
 "execute :post-groups with {"name" "0", "path" "00"}"
 "when the query op. is executed again, the response is NOT equal 
  to S"]

["State S = response from the given query operation"
 "execute :delete-groups with {"id" 0}"
 "execute :post-groups with {"name" "0", "path" "00"}"
 "when the query op. is executed again, the response is NOT equal 
  to S"]

["State S = response from the given query operation"
 "execute :get-groups with {}"
 "execute :post-groups with {"name" "0", "path" "00"}"
 "when the query op. is executed again, the response is NOT equal 
  to S"]

;; state mutation with increase, :post-groups as query operation
"no novel example found"
\end{minted}
    \caption{Exploration result of GitLab with MP-S-4.}
    \label{fig:gitlab-MPS4}
\end{figure}

Examples of MP-S-4 (State Mutation with state size increase) are found in Figure~\ref{fig:gitlab-MPS4}, both for \texttt{:get-groups} and \texttt{:post-groups} as given query candidates. As before, we see the get-post-get sequence, and for examples that must include a \texttt{:get-groups} (L13-17) or \texttt{:delete-groups} (L7-11) in the sequence, the same pattern appears with the addition of the required extra operation. As we can note in lines 19-20, no sequence was found that satisfied this property with \texttt{:post-groups} as the query operation candidate. These results indicate that MP-S-4 does a good job of verifying query-like behaviour.

MP-S-5, where the size of the state should decrease, will result in the same type of sequences as MP-S-3 (post-get-delete), but also result in the same benefit as MP-S-4; it gives no false positives, in our exploration, given a non-query operation as input.

\subsubsection{Exploration Summary}

The presented evaluation of the proposed meta-properties indicates that the response-based properties are strong in providing minimal, relevant examples of both query-like and create-like behaviour. In addition, these properties correctly show examples of boundaries where operation parameters result in different behaviours.
The state-based meta-properties are more varied in their usefulness. To further verify the findings of the response-based properties, our evaluation strongly suggests that not only a change in the state should be checked, but also the \textit{size} of the change.

\begin{result}
Relevant examples of query- and create- operations can be found by our approach, using MP-R-1 and MP-R-2, and further verified with state mutation properties which consider sizes, MP-S-4 and MP-S-5. Other proposed meta-properties provide no additional relevant examples and can include false positives. \vspace{1em}

We have shown that by using general meta-properties we can find relevant short examples, and doing so without the need for a formal specification or access to source code. We achieve this by generating abstract example sequences, with parameters and references, and select and shrink sequences that conform to the sought general behaviour.
\end{result}

We end this section with an experience from our evaluation to further show how we interactively learnt more about the API under exploration. While experimenting on GitLab, without the soft-reset enabled, we got inconsistent results. Some properties we expected to give examples did not. It turned out that, as many industry-grade APIs do, GitLab implements a paging mechanism for its query operations. The results we got were due to the fact that when more than 20 entities existed, only the first 20 were returned, resulting in no apparent state change when an entity had been created. These experiments taught us about the paging mechanism of the API. The solution was to enrich the AMOS with the information that the get operation is a ranged operation. The \emph{API Translation} component for a REST API would then interpret this information and perform multiple calls, returning the complete state to the example generation process. In terms of internals of the method, the get operation was still one operation with one response, only the translation component needed to be adapted.

\section{Related Work}\label{sec:related-work}

Much work has been done in the area of producing examples to help developers understand an API, by different means of automation \cite{Buse-Synthesizing-API-usage-examples-2012, Barnaby-Exempla-Gratis-(E.G.)-Code-Examples-for-Free-2020, Gu-CodeKernal-2019,Kim-Adding-Examples-into-Java-Documents-2009, Mar-Recommending-Proper-API-Code-Examples-for-Documentation-Purpose-2011, Montandon-Documenting-APIs-with-examples-Lessons-learned-with-the-APIMiner-platform-2013, Mittal-Generating-examples-for-use-in-tutorial-explanations-using-a-subsumption-based-classifier-1994, Gerdes-Understanding-Formal-Specifications-through-Good-Examples-2018, Holmes-Approximate-Structural-Context-Matching:-An-Approach-to-Recommend-Relevant-Examples-2006, Moreno-How-Can-I-Use-This-Method-2015}. The most common approach has been to rely on white-box information, a corpus of source code example uses of the API~\cite{Buse-Synthesizing-API-usage-examples-2012, Barnaby-Exempla-Gratis-(E.G.)-Code-Examples-for-Free-2020, Gu-CodeKernal-2019, Kim-Adding-Examples-into-Java-Documents-2009, Mar-Recommending-Proper-API-Code-Examples-for-Documentation-Purpose-2011, Montandon-Documenting-APIs-with-examples-Lessons-learned-with-the-APIMiner-platform-2013, Holmes-Approximate-Structural-Context-Matching:-An-Approach-to-Recommend-Relevant-Examples-2006, Moreno-How-Can-I-Use-This-Method-2015}, which we do not. 

In addition to fully automated approaches, Head et al. propose an interactive and iterative approach where the user is involved in selecting code examples from their own code \cite{Head-Interactive-Extraction-of-Examples-from-Existing-Code-2018}. As demonstrated, we also base our method on an interaction with the user, where knowledge gained from one set of properties are input to the next set of properties. But again, we do not require access to the code of the system. 

Instead of only generating usage examples as API interactions, examples in natural language can be produced, based on different kinds of specifications \cite{Lavoie-The-ModelExplainer-1996, Swartout-GIST-English-Generator-1982, Burke-Translating-Formal-Software-Specifications-to-Natural-Language-2005}. In this work, we make an initial attempt to ``humanize'' the operation interactions of generated examples, in natural language, while still abstracting away from the details of the SUT. However, how users prefer to be presented with generated examples remains to be investigated.

A different approach, from using existing example usages of an API as a source, is introduced by Mittal et al. In this approach, examples of Lisp syntax are generated, given a Lisp syntax grammar \cite{Mittal-Generating-examples-for-use-in-tutorial-explanations-using-a-subsumption-based-classifier-1994}. As with our method, Mittal et al. have a strong focus on interesting examples. However, their scope is limited to Lisp, whereas we aim to be more generally applicable, abstracting the SUT. 

The most similar work to ours is the method proposed by Gerdes et al. \cite{Gerdes-Understanding-Formal-Specifications-through-Good-Examples-2018}. Gerdes et al. presented heuristics for choosing examples for a stateful API, and showed that subjects shown those examples were better able to predict the API's behaviour than subjects shown examples selected to cover all the code. When generating test cases, code coverage is a common metric for selection, but Gerdes at al. showed that different criteria are needed to select relevant examples. As Gerdes et al. do, we also base our method on finding examples in a black-box fashion with the use of a test generation technique with shrinking (property-based testing). However, the major difference in our methods is that while Gerdes et al. require a formal specification of the behaviour of the SUT, we do not. Instead, we only require a specification of available operations and their inputs, with no behavioural information. Our main novelty is the use of a set of general meta-properties to explore and generate examples of the behaviour of the SUT.


\section{Conclusions}\label{sec:conclusions}

Automatically generated examples can help developers to understand an API. In addition, exploring the behaviour of an API can reveal unwanted behaviours or validate expected behaviour.
In this paper, we have proposed a novel method to generate relevant examples of API behaviours. We do so by using test generation, searching for examples of behaviours. The behaviours are based on meta-properties, abstracting the behaviour of a specific API in a general way.
Our evaluation, on an industry-grade REST API, shows the applicability of the method in finding good, relevant, and small examples of behaviour. We also show how different meta-properties provide new knowledge, which can aid developers and users in understanding their systems, without the labour of producing a formal specification.

We identify some areas in which this work can be extended in future work.
One such area is to define and assess additional meta-properties, to further explore a system. More studies would be needed to assess how additional general meta-properties could be used on different kinds of systems.

Another area of future work is the reporting of results. As we have mentioned, producing different kinds of reports are an engineering effort. However, to know how users best absorb the information in the report and what they want to be included, and if this is different for different kinds of examples, would require further study. For example, should examples be presented in a more visually appealing manner, or as a report including natural text, or in some domain language relevant to the domain under exploration? 

Finally, how users best use, deploy, and integrate a system exploration approach in their workflow is an open question. As Gerdes et al.~\cite{Gerdes-Understanding-Formal-Specifications-through-Good-Examples-2018} showed, users presented generated examples were better able to predict the behaviour of the system, but how can we best help the users to interactively explore the system by themselves---extracting as much knowledge about the systems behaviour with as little effort as possible.

We introduced this paper with the question: \emph{Can we generate tests to explore the behaviour of the system without access to a formal specification or source code?} The answer is yes! We have presented an approach that does not rely on any inputs except for general behaviours and have shown that this indeed suffices to enable the automatic generation of relevant examples that allow us to explore a system's behaviour.

\section*{Acknowledgments}
This work is supported by ABB, the industrial postgraduate school Automation Region Research Academy (ARRAY) funded by The Knowledge Foundation (KKS).

\appendix

\begin{figure}[h]
  \centering
\begin{minted}{clj}
(def persons-db (atom {}))

(defn post-person
  [db person]
  (swap! db assoc (:person/name person) person))

(defn get-persons
  [db]
  (if (seq @db)
    (vec (vals @db))
    []))

(defn delete-person
  [db name]
  (swap! db dissoc name))
\end{minted}
    \caption{An example application with a simple API.}
    \label{fig:clojure-app}
\end{figure}

\begin{figure}[h]
  \centering
\begin{minted}{c}
persons_db = []

post_person (person p) {
  persons_db[p.name] = p
}

get_persons () {
  if (not-empty persons_db)
    return persons in persons_db
  else 
    return []
}

delete_person (string name) {
  delete person_db[name]
}
\end{minted}
    \caption{Pseudo-code of example application with a simple API.}
    \label{fig:pseudo-app}
\end{figure}

\section{Code Walkthrough}\label{app:code-walkthrough}

In this appendix, we provide a code walkthrough of the running example application presented in Section~\ref{sec:running-example}. The running example is written in Clojure~\cite{Hickey-Clojure-2020} (see Figure~\ref{fig:clojure-app}), and pseudocode of the same example is provided (in Figure~\ref{fig:pseudo-app}), for the reader unfamiliar with Clojure.

The code in Figure \ref{fig:clojure-app} starts by defining the in-memory state, using the \texttt{atom} function, named \texttt{persons-db}. The \texttt{persons-db} is initialised as an empty hash-map, with the \texttt{\{\}} literal. 

The \texttt{post-person} function uses the \texttt{swap!} function to update the current state of the database. The second argument to \texttt{swap!} is a function that produces the new value of the database. In this case, we \texttt{assoc}iate the current state of the database with a new hash-key and value. The name field of the person is the hash-key, and the value for this hash-key is the person entity given. Note that there is no check of whether a user with a given name already exists---it will just be overwritten if so. This is a behaviour for which our method can produce an example and is something we will address when we explore the behaviour of this application in Section \ref{section:an-exploration-example}.

The function \texttt{get-persons} returns the values stored in the database. The current state of the database is read with the \texttt{@} literal. The database is stored as a hash-map with names as keys and the actual person entities as values. In the case of \texttt{get-persons}, we only want to return the entities, not the index keys. Therefore, we use the \texttt{vals} function to extract only the values of the hash-map. Finally, we put the values in a \texttt{vec}tor. If there is no \texttt{seq}uence of values in the database, we return an empty vector.

The last function, \texttt{delete-person}, also uses the \texttt{swap!} function to update the current state of the database. In this case, we \texttt{dissoc}iate the key in the hash-map corresponding to the given name, removing this entry from the database.

\newpage

\bibliographystyle{acm}
\bibliography{main}

\end{document}